\documentclass[12pt]{iopart}
\usepackage{graphicx}
\begin{document}

\title{Unconventional spin texture of a topologically nontrivial semimetal Sb(110)}

\author{A Str\'{o}\.{z}ecka$^1$, A Eiguren$^{2,3}$, M Bianchi$^4$, D Guan$^4$, C H Voetmann$^4$, S Bao$^5$, Ph Hofmann$^4$ and J I Pascual$^1$}
\address{$^1$ Institut f\"{u}r Experimentalphysik, Freie Universit\"at Berlin, 14195 Berlin, Germany}
\address{$^2$ Departameto de F\'isica de la Materia Condensada, EHU/UPV, Barrio Sarriena sn 48940 Leioa,
Spain.}
\address{$^3$ Donostia International Physics Center (DIPC), Paseo Manuel de Lardizabal, 4. 20018 Donostia-San Sebastian, Spain.}
\address{$^4$ Department of Physics and Astronomy, Interdisciplinary Nanoscience Center, Aarhus University,
8000 Aarhus C, Denmark}
\address{$^5$ Department of Physics, Zhejiang University, Hangzhou, 310027 China}

\date{\today}

\begin{abstract}
The surfaces of antimony are characterized by the presence of spin-split
states within the projected bulk band gap and the Fermi contour is thus
expected to exhibit a spin texture. Using spin-resolved density functional
theory calculations, we determine the spin polarization of the surface
bands of Sb(110). The existence of the unconventional spin texture is
corroborated by the investigations of the electron scattering on this
surface. The charge interference patterns formed around single scattering
impurities, imaged by scanning tunneling microscopy, reveal the absence of
direct backscattering signal. We identify the allowed scattering vectors
and analyze their bias evolution in relation to the surface-state
dispersion.
\end{abstract}


\maketitle

Pure bismuth and antimony are group-V semimetals and their surfaces support spin-orbit split surface states existing within the projected bulk band gap %\cite{Agergaard2001,Hofmann2006,Bianchi:2012,Sugawara:2006,kadono:2008}
\cite{Agergaard2001}- \cite{kadono:2008}. The two materials differ
however, with respect to the topological character of their bulk bands.
Whereas pure bismuth is topologically trivial, adding a small amount of Sb
into Bi is enough to drive the system into a topological insulator phase
\cite{Hsieh2008,Guo:2011}. The spin-nondegenerate surface states of the
topological insulator Bi$_{(1-x)}$Sb$_x$ (0.09$<$x$<$0.18) continuously
connect the valence and the conduction band and their existence is derived
directly from the fundamental considerations of the parity characteristics
of the bulk bands \cite{Teo:2008}. As more Sb is added into the alloy, the
system becomes semimetallic again, it keeps however the nontrivial
topological order of the bulk bands \cite{Guo:2011}. Thus, the other
extreme case of the Bi$_{(1-x)}$Sb$_x$ alloy, pure antimony, is a
semimetal with bulk characteristic of a strong topological insulator.

The (111) surfaces of the topologically nontrivial materials, including the Bi$_{(1-x)}$Sb$_x$ alloy and Sb, have been extensively investigated by spin and angle resolved photoemission spectroscopy (ARPES) and scanning tunneling microscopy (STM) %\cite{Sugawara:2006,kadono:2008,Hsieh2008,Guo:2011,Hsieh2009b,Hsieh:2010,Park:2011,Roushan2009}
\cite{Sugawara:2006}-\cite{Guo:2011},\cite{Hsieh2009b}-\cite{Roushan2009}.
These studies allowed to determine the spin textures
\cite{kadono:2008,Hsieh2009b,Hsieh:2010}, map the surface band dispersion
\cite{Sugawara:2006}-\cite{Guo:2011} and revealed an unconventional
electron dynamics \cite{Park:2011,Roushan2009,Gomes:arxiv}. At the same
time, studies addressing non-(111) surfaces of those materials are almost
nonexistent. Only recently, we have shown that Sb(110) exhibits spin-split
surface states but, contrary to predictions based on its bulk topology,
their dispersion properties are those of a trivial material . This is due
to the semimetallic character of the surface, which relaxes the character
of the surface state bands imposed by the bulk topology
\cite{Bianchi:2012}. The spin polarization of these surface bands remains,
however, still unknown. Since Antimony has a weaker spin-orbit coupling
than bismuth, it is crucial to prove if a sufficient degree of spin
polarization still remains in the split surface bands and how it affects
its electron scattering properties.

The spin texture of the surface bands can be addressed indirectly by investigating the electron scattering processes %\cite{Roushan2009,Pascual2004,Alpichshev2010,Zhang2009a,Strozecka:2011}
\cite{Roushan2009},\cite{Pascual2004}-\cite{Strozecka:2011}.  The
spin-orbit coupling  affects the electron scattering probabilities, making
them dependent on the overlap of the spin wavefunctions of the final and
initial scattering state. In particular, the characteristic feature of
spin-orbit coupled systems is the suppression of direct backscattering,
because the states with opposite momenta are required to have orthogonal
spins by time reversal symmetry \cite{Pascual2004,Petersen:2000}. The
modulation in the charge local density of states (LDOS) which arises due
to the interference of the electrons scattering at surface impurities is
directly accessible with scanning tunneling microscopy and spectroscopy
(STS). The scattering vectors $q$, connecting the initial and final states
in the reciprocal space, can be identified by examining the Fourier
transform (FT) of the interference patterns \cite{Petersen1998}. By
demonstrating the lack of backscattering signal, this method allowed to
confirm the existence of the spin texture at the surfaces of semimetallic
bismuth and antimony \cite{Gomes:arxiv,Pascual2004,Strozecka:2011} as well
as topological insulators Bi$_{(1-x)}$Sb$_x$ , Bi$_2$Se$_3$ and
Bi$_2$Te$_3$ \cite{Roushan2009,Alpichshev2010,Zhang2009a}.

In this paper, we investigate the electron scattering processes on a
non-(111) surface of topologically nontrivial material, Sb(110), in
order to gain experimental insight into the spin texture of its
surface electronic bands. We observe strong modulation in the LDOS
around single adatoms on the surface and correlate the features in
the Fourier transformed images to the specific scattering vectors. We
identify those scattering events by analyzing the Fermi contour
measured by ARPES and the spin polarization of the surface bands
predicted by relativistic density functional theory (DFT)
calculations. The resulting spin texture is highly anisotropic and
strongly varies with the character of the bands, deviating from the
conventional picture of a Rashba surface. The experimental STM data
do not show any direct backscattering between the nondegenerate
surface bands, confirming their spin-split nature and the spin
texture of the Fermi contour.

We have investigated the surface electronic structure of Sb(110) using STM
and ARPES. The STM exeriments were performed in a custom-made microscope
working in an ultrahigh vacuum at 5\,K. The ARPES data were taken at the
SGM-III beamline of the synchrotron radiation facility ASTRID in Aarhus
\cite{Hoffmann:2004}. The combined energy  and angular resolution were
better than 10 meV and  $0.13^\circ$, respectively. The data shown here
were all taken using a photon energy of 20~eV. During the photoemission
measurements the sample was kept at a temperature of 60~K. The Sb(110)
single crystal surface was cleaned $in situ$ by repeated sputtering and
annealing cycles.

We modeled the electronic and spin structure of the Sb(110) using a
non-collinear DFT \cite{Corso:2005,pwscf} within the
Perdew-Burke-Ernzerhof implementation \cite{Perdew:1996} of the
generalized gradient approximation. We considered a repeated slab system
consisting of 54 layers relaxed up to forces $< 10^{-4}$~Ry/a.u. and used
fully relativistic norm-conserving pseudopotentials as described in Ref.
\cite{Corso:2005} with the energy cutoff corresponding to E$_c$=60~Ry. In
order to calculate the projection of the bulk band structure onto the
(110) surface, we have also used the tight-binding scheme of Liu and Allen
\cite{Liu:1995}.

\begin{figure}
\includegraphics[width=0.9\columnwidth]{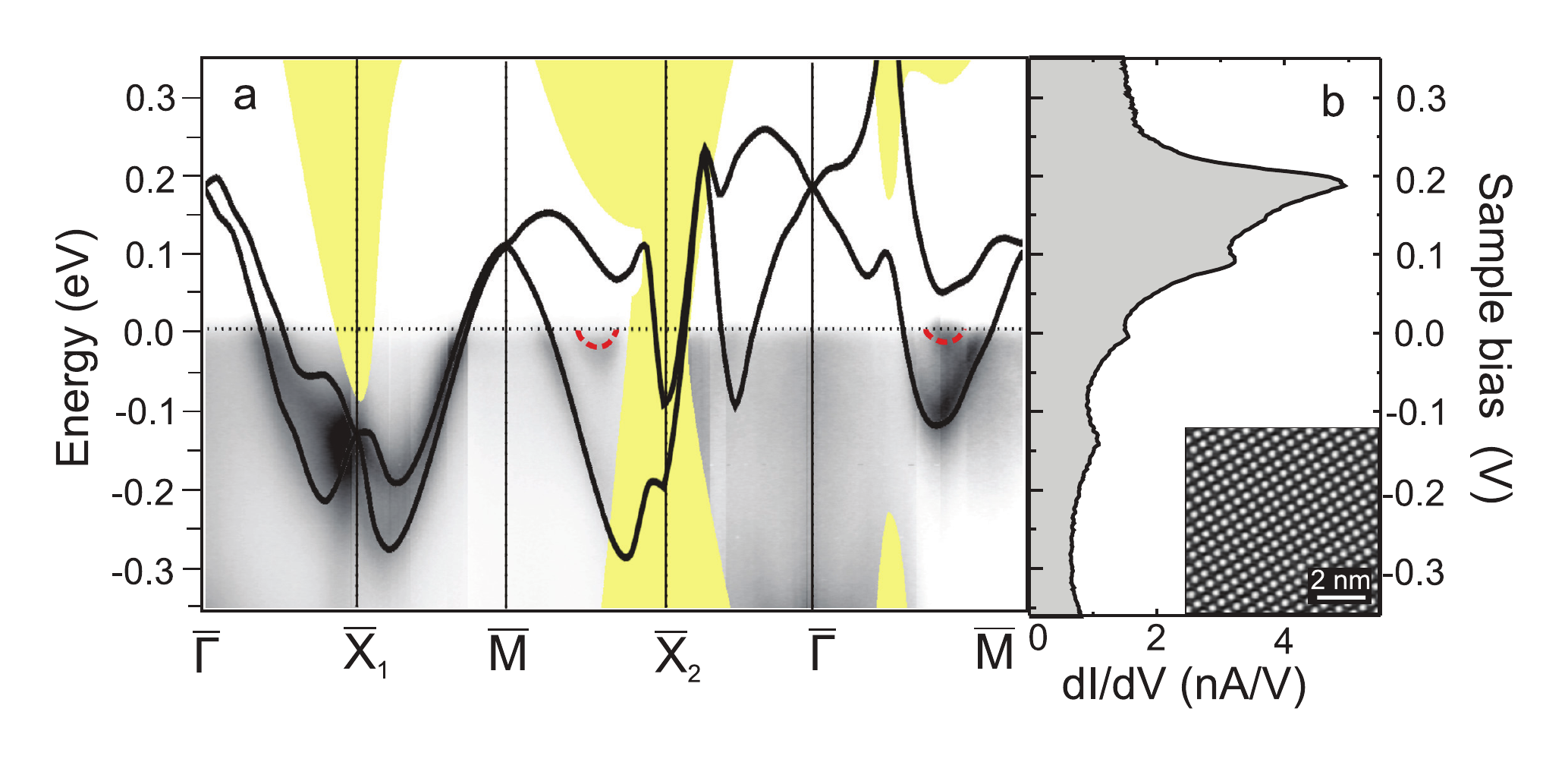}
\caption {a) Surface state dispersion of Sb(110) measured by ARPES and calculated
by DFT. The ARPES intensity is coded in the grey scale; the continuous lines
correspond to the calculated surface states. Yellow shading depicts the projected
bulk states. The dashed line highlights the shallow electron pockets revealed by
ARPES, but missed in DFT. (b) dI/dV spectrum of a clean Sb(110) surface, showing the LDOS in the bias range of the surface states. The features in the spectrum correspond to the onset of the surface bands. The inset image shows the atomically resolved topography of the clean surface.}\label{fig:fig1ARPES}
\end{figure}

\begin{figure}
\includegraphics[width=0.9\columnwidth]{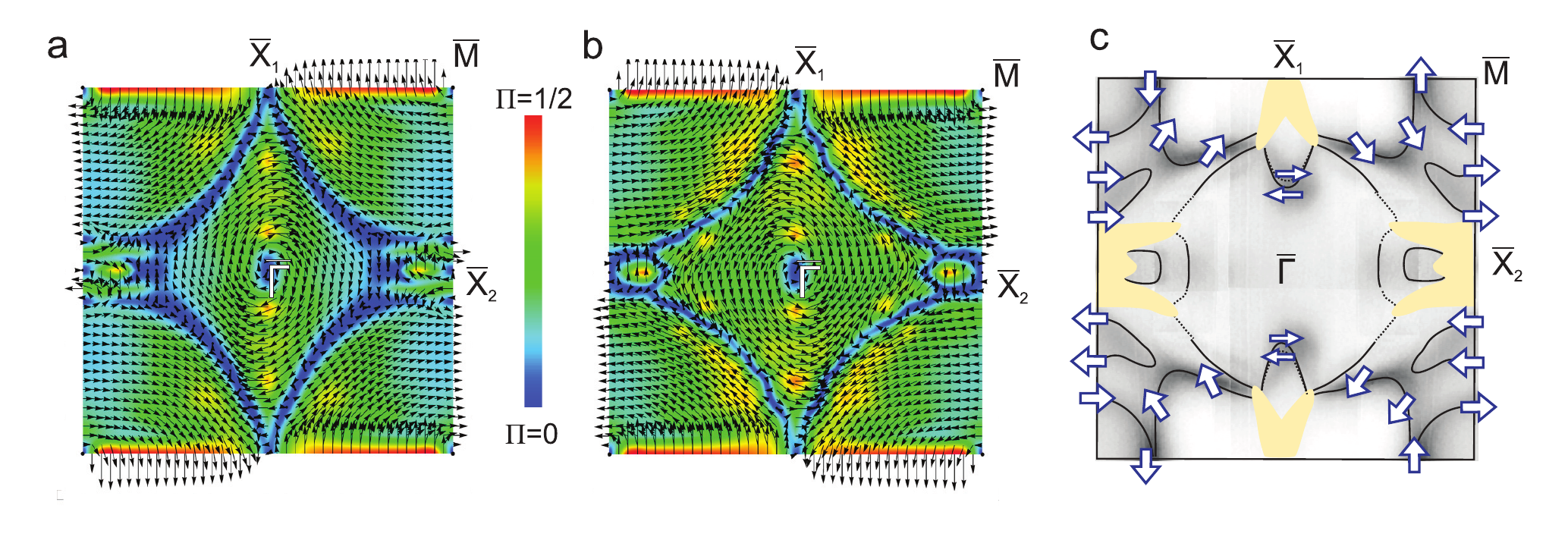}
\caption {Calculated spin texture within first surface Brillouin zone, for
the lower (a) and upper (b) subband. The color code
represents the degree of the spin polarization $\Pi$ (see text for
details), arrows show direction of the spin in the surface
plane. (c) Fermi contour of Sb(110) measured by ARPES. The continuous lines are the schematic representation of the ARPES data; the dotted line represents parts of the contour with vanishing ARPES intensity. The arrows schematically show the spin polarization of the contour, as deduced from plots in (a) and (b).}\label{fig:fig1SPIN}
\end{figure}

The Sb(110) surface shows several spin-orbit split surface states
within the projected bulk band gap. In Figs.\,\ref{fig:fig1ARPES} and
\ref{fig:fig1SPIN} we present electronic structure of Sb(110) as
measured by ARPES and calculated by DFT. The dispersion of the
surface bands and their topological character were discussed in
detail in Ref.\,\cite{Bianchi:2012}. Here, we summarize the most
important properties of this surface, which directly affect the
electron scattering processes, i.e. the dispersion of the surface
bands in the vicinity the Fermi level and their spin texture. The
electronic structure of Sb(110) along certain high symmetry
directions is shown in Fig.\,\ref{fig:fig1ARPES}(a) and the
experimentally determined Fermi contour is presented in
Fig.\,\ref{fig:fig1SPIN}(c) (grey-scale background). For clarity, the
lines have been superimposed on the data in
Fig.\,\ref{fig:fig1SPIN}(c), showing schematically the shape of the
contour. The most important spectroscopic features which we identify
in both calculations and the experimental data are: (i) the hole
pocket around the $\bar{M}$ point, giving rise to circular contour in
Fig.\,\ref{fig:fig1SPIN}(c), (ii) a pocket around $\bar{X}_1$,
identified in the Fermi contour as a 'butterfly' shaped feature;
(iii) an extended hole pocket around $\bar{\Gamma}$, observed as a
weak feature in ARPES data. In addition, ARPES reveals a very shallow
electron pocket (dotted lines in Fig.\,\ref{fig:fig1ARPES}(a) in
$\bar{X}_2 \bar{M}$ and $\bar{\Gamma} \bar{M}$ directions), which is
not well reproduced by the calculations.
%These discrepancies between experiment and theory can be easily accounted for by the small uncertainties in the band dispersion and the Fermi level position in the calculations [].
This pocket is particularly important to understand the electron
scattering dynamics in the vicinity of the Fermi level, since it
contributes strongly to the density of states in this energy window
\cite{Pascual2004}. In fact, the experimentally measured dI/dV
spectrum (Fig.\,\ref{fig:fig1ARPES}(b)), which reflects the LDOS of a
clean Sb(110) surface, shows a peak in the vicinity of the Fermi
level, marking the onset of the electron pocket. Several other sharp
features in the dI/dV spectrum correspond to the onset (negative
bias) and the top (positive bias) of the surface bands.
%The presence of rather narrow features in the LDOS spectrum Sb(110) is one of the consequences of a relatively small spin orbit coupling, which focuses the spin-split states in a narrow energy window.
Due to the relatively weak spin-orbit coupling in Sb, the splitting of the
bands is rather small so that in certain parts of the Brillouin zone the
two spin partners are close to being degenerate. Accordingly, some of the
features of the Fermi contour in Fig.\,\ref{fig:fig1SPIN}(c), the high
intensity feature along $\bar{\Gamma} \bar{X}_1$ direction and crossing
along $\bar{X}_1 \bar{M}$ direction, are double crossings and the two
states with opposite spin become almost degenerate.

To determine the spin of the Fermi contour we calculated the spin texture
of the upper and lower subband within the first Brillouin zone, as shown
in Fig,\,\ref{fig:fig1SPIN}(a) and (b), respectively. The degree of
certainty about the electron spin, $\Pi$ $\equiv$ $\sqrt{\sum_i \langle
S_i \rangle^2}$,
is represented by the color scale in the background ($\Pi$ $\in$ $[0,1/2]$). The direction of the spin in the surface plane is depicted by arrows. %The spin in plane is ill-defined near the $\bar{X}_2$ point, where the surface bands enter the bulk bands and become surface resonances.
The polarization out of plane becomes important only for the lower
subband, close to the $\bar{X}_2$ point. Across the rest of the Brillouin
zone, the expectation value of the polarization out of plane remains very
small. The spin texture of any constant energy contour can be now defined
by considering separately the spin polarization for the lower and upper
subband. In Fig.\,\ref{fig:fig1SPIN}(c) the spin polarization deduced from
plots in (a) and (b) is superimposed on the Fermi contour of Sb(110). We
find that the spin direction is almost constant as we move along one side
of the butterfly-shaped pocket and rotates along the $\bar{M}$ pockets.
The features along $\bar{\Gamma} \bar{X}_1$ and  $\bar{X}_1 \bar{M}$
correspond to almost double crossings and, consequently, have
contributions from both spin-split bands, and thus, two opposite spins.

\begin{figure}
\includegraphics[width=0.9\columnwidth]{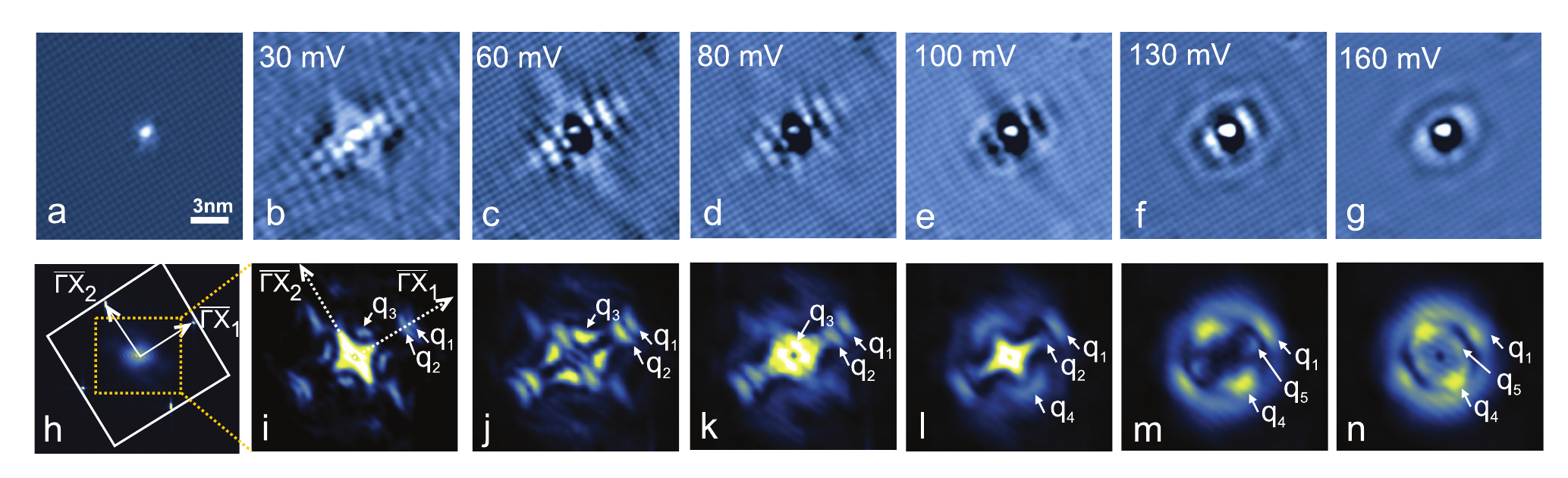}
\caption{STM imaging of the interference patterns formed around a single
impurity atom on Sb(110). The adatoms (presumably Sb) have been obtained
by control tip indentation in the substrate.  (a) Constant current
topography image of the impurity (I=0.5\,nA, V=60\,mV). (b)-(g) Energy
resolved dI/dV images showing electron interference patterns, at bias
voltages 30\,mV, 60\,mV, 80\,mV, 100\,mV, 130\,mV and 160\,mV,
respectively. A corresponding FFT image is shown below each real space
image: (h)\,-\,(n). The full FFT of the topography image (h) shows the
main crystal directions and bright points corresponding to the atomic
corrugation. The images (i)\,-\,(n) have been cut out of the full FFT
(dotted square in (h)) for a more detailed view of the scattering vectors.
The scattering vectors referred to in the text are marked as
$q_1$-$q_5$.}\label{fig:fig2}
\end{figure}

The spin texture of the constant energy contours will affect the
scattering of the electrons by the surface impurities. In particular, the
systems with spin-orbit split surface states are characterized by the
absence of direct backscattering, imposed by the time reversal symmetry
\cite{Petersen:2000}. In order to experimentally verify the spin texture
of Sb(110) and identify the allowed scattering vectors we analyzed the
charge interference patterns formed by electrons scattering on a single
adatom (Fig.\,\ref{fig:fig2}). In the bias range between -80 and 240\,meV
we observe pronounced modulation in the LDOS signal. A series of dI/dV
spectroscopic images in Fig.\,\ref{fig:fig2}(b)-(g) shows the evolution of
the patterns for several chosen bias voltages. The Fast Fourier Transform
(FFT) of the images ((i)-(n)) shows high intensity features, corresponding
to the allowed scattering vectors $\overrightarrow{q}$.

\begin{figure}
\includegraphics[width=0.9\columnwidth]{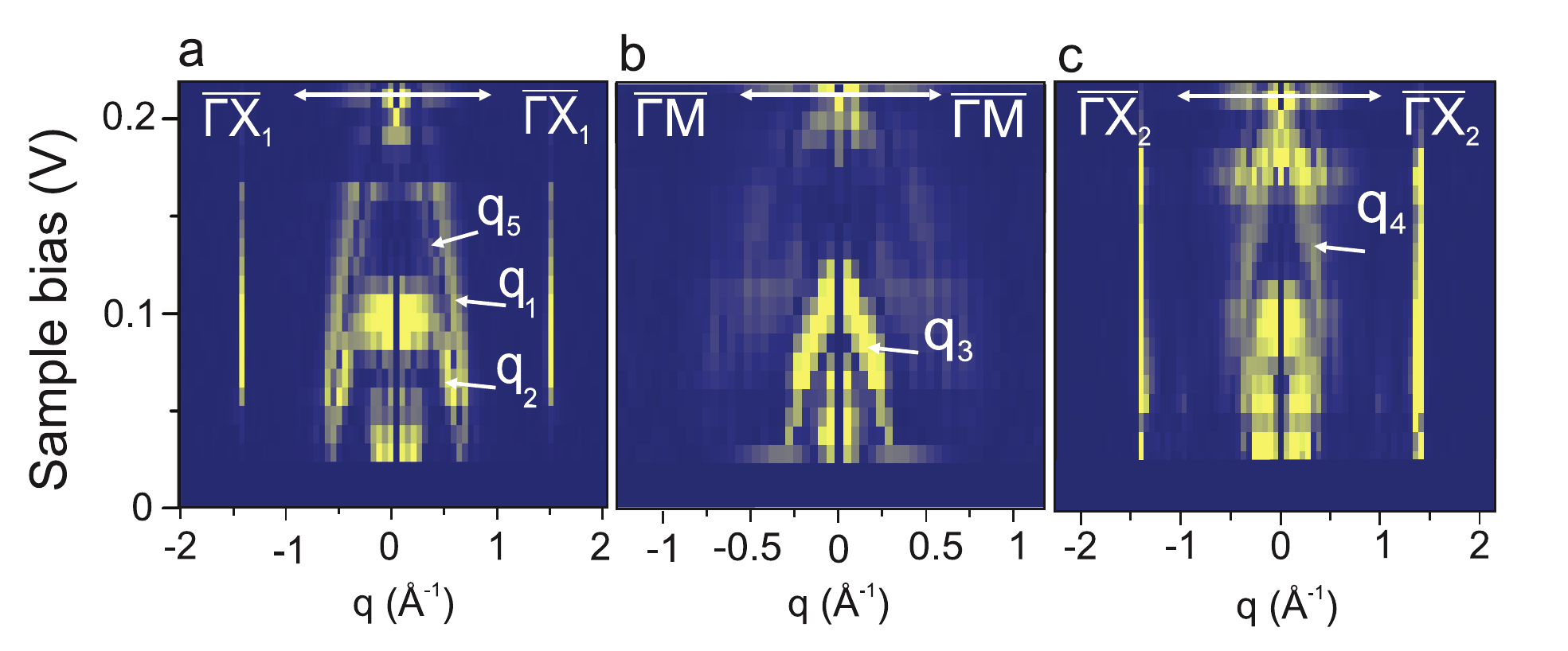}
\caption {Dispersion of the scattering vectors in the directions $\bar{\Gamma} \bar{X}_1$ (a), $\bar{\Gamma} \bar{M}$ (b) and $\bar{\Gamma} \bar{X}_2$ (c). The non-dispersive points correspond to the the atomic corrugation spots in the FFT. The labelling of the scattering vectors is the same as in Fig.\,\ref{fig:fig2}.}\label{fig:fig3}
\end{figure}

Close to the Fermi level ((i) and (j)) the most distinctive features of
the FFT \cite{note} are two double arc-shaped features along $\bar{\Gamma}
\bar{X}_1$ direction (labeled $q_1$ and $q_2$) and the four spots lying
symmetrically around the center (labelled $q_3$). The features disperse
with increasing sample bias as shown in more detail in
Fig.\,\ref{fig:fig3}(a)-(c). Each graph represents a set of the linescans
through the FFT images taken at different sample bias in certain high
symmetry direction ($\bar{\Gamma} \bar{X}_1$, $\bar{\Gamma} \bar{M}$ and
$\bar{\Gamma} \bar{X}_2$, respectively). The vectors $q_1$ and $q_2$ get
smaller until at around 100\,meV the feature $q_2$ vanishes from the FFT
and the $q_1$ transform into a cicular shape. The four features $q_3$ move
to the center of the FFT and disappear also around at 100\,meV. At the
same bias range two more scattering events appear: the features $q_4$ in
the direction $\bar{\Gamma} \bar{X}_2$ and $q_5$ in the direction
$\bar{\Gamma} \bar{X}_1$.

In order to establish the origin of the features observed in FFT, we
consider the shape and spin texture of the constant energy contour. In the
simple picture, the scattering probability  $P(\vec{q},eV)$ at energy
$eV$, between the $\overrightarrow{k}$-states separated by the scattering
vector $\overrightarrow{q}$, is determined by the electron density of
states $\rho$ of the final and the initial state, weighted by the overlap
of the spinor wavefunctions $\varphi_S$ \cite{Roushan2009}:
\begin{eqnarray} % \label{eq1}
P(\vec{q},eV)= \int \rho(\vec{k},eV) \rho(\vec{k}+\vec{q},eV) \langle \varphi_S (\vec{k}) | \varphi_S (\vec{k}+\vec{q}) \rangle d^2k
\end{eqnarray}
Thus, the spin dependent part of the scattering amplitude will be maximal
if the initial and final spins are parallel and zero when the two spins
are antipararell.

If the spin contribution to the scattering amplitudes is neglected, the
map of the scattering vectors allowed by the Fermi contour can be
evaluated directly from the photoemission data, by calculating the
convolution of the ARPES spectra. The autocorrelation image (AC-ARPES) can
be derived from the ARPES intensities $I(\vec{k},eV)$, as
\cite{Markiewicz:2004,McElroy2006}:
% The JDOS picture can be obtained from the single particle spectral function $A(\vec{k},eV)$ measured by ARPES, by calculating the autocorrelation image \cite{Roushan2009,Markiewicz:2004,McElroy2003,McElroy2006}:
 \begin{eqnarray} %\label{eq2}
AC-ARPES(\overrightarrow{q},eV)=\int I(\overrightarrow{k},eV) I(\overrightarrow{k}+\overrightarrow{q},eV) d^2k.
 \end{eqnarray}

The AC-ARPES defined in this way reproduces the features of the joint
density of states (JDOS), which in turn maps out the LDOS contribution to
the scattering probabilities. While the quantitative comparison between
AC-ARPES and JDOS is not straightforward, since the ARPES intensities $I$
are related to the electron spectral function through an energy and
$k$-dependent matrix element, the autocorrelation image stills allows for
a qualitative analysis of the scattering vectors
\cite{Markiewicz:2004,McElroy2006}. For spin degenerate Fermi contour, the
AC-ARPES can be directly compared to the FFT of the interference patterns
resolved by STM at low bias \cite{McElroy2006}.

\begin{figure}
\includegraphics[width=0.9\columnwidth]{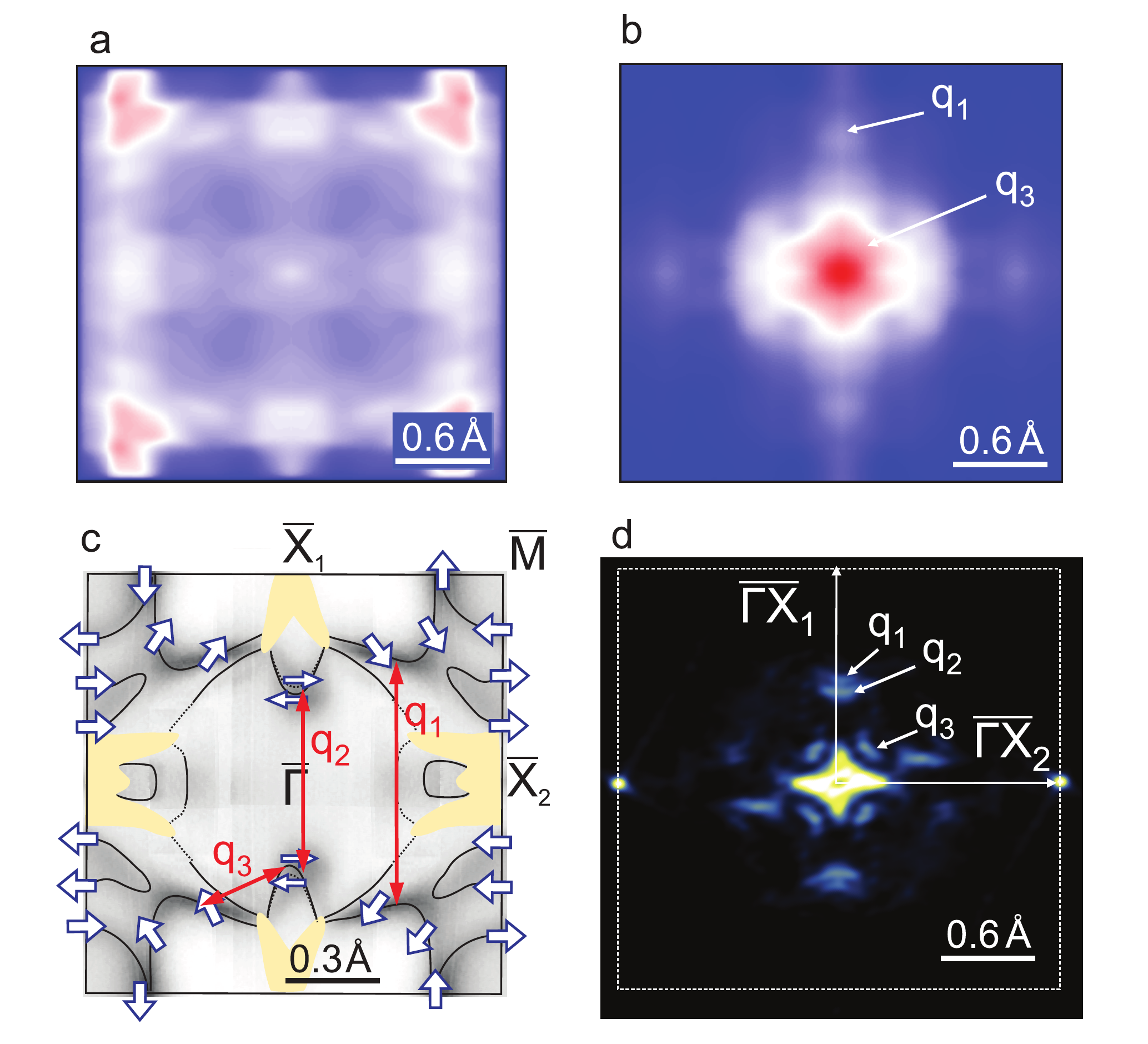}
\caption {(a) The autocorrelation of the ARPES data of the Fermi contour
of Sb(110), neglecting its spin texture. (b) The autocorrelation
signal as in (a), but with suppressed backscattering (see text for
details). (c) The Fermi contour of Sb(110) with identified scattering
vectors. (d) FFT measureded at bias close to the Fermi level (30mV).   }\label{fig:fig4}
\end{figure}

We have calculated the AC-ARPES map for Sb(110) (Fig.\,\ref{fig:fig4}(a)),
neglecting its spin texture, and compared it to the FT-STM map measured
close to the Fermi level (Fig.\,\ref{fig:fig4}(d)). We find that certain
parts of the AC-ARPES are suppressed in the STM data. In fact, many of the
features present in AC-ARPES but absent in the FFT map can be identified
as direct backscattering events, which are expected to be prohibited by
the spin texture of Sb(110). Thus, to describe the experimental STM data
more accurately, we introduce into the cross-correlation integral
(Eq.\,(2)) an additional term:
$1-\cos[\angle(\overrightarrow{k},\overrightarrow{k}+\overrightarrow{q})]$.
While this basic term does not account for the anisotropic spin texture of
Sb(110), it suppresses the  backscattering in a simple way. The result
(Fig.\,\ref{fig:fig4}(b)) reproduces the FFT map to a greater degree.
Large parts of the original AC-ARPES map in Fig.\,\ref{fig:fig4}(b) are
now diminished and the remaining features correspond relatively well to
the main spots of the FFT. We can thus conclude that the backscattering is
indeed absent on Sb(110) surface, confirming the spin-split character of
its surface bands.

We can now associate the spots resolved in the FT-STM map
(Fig.\,\ref{fig:fig4}(d)) with the specific scattering vectors.
%The events of the highest scattering probability will involve the parts of the Fermi contour with the highest DOS. Those we can  the shallow electron pocket in $\bar{\Gamma} \bar{M}$ direction (appears in Fermi contour as the edge of the butterfly wing).
First, we identify the feature $q_1$ as interband scattering between
the edges of the 'butterfly wings' in  $\bar{\Gamma} \bar{X}_1$
direction (Fig.\,\ref{fig:fig4}(c)). This feature of the Fermi
contour corresponds to a very shallow and extended electron pocket
and is thus expected to have a high DOS and contribute strongly to
the scattering at this energy range. With increasing bias, $q_1$
disperses very steeply (see Fig.\,\ref{fig:fig3}(a)), reflecting the
dispersion of the left band which forms the electron pocket in
$\bar{\Gamma} \bar{M}$ direction. In principle, the scattering
between the 'butterfly wings' should also give rise to features in
FFT in the  $\bar{\Gamma} \bar{X}_2$ direction. They are indeed
resolved in the AC-ARPES images Fig.\,\ref{fig:fig4}(a) but we
observe them only as faint spots in FT-STM image, possibly due to
partial suppression by the overlap of the spinor wavefunctions.
Next, we assign the feature $q_2$ to the backscattering between the $\bar{\Gamma} \bar{X}_1$-crossing points. At first glance, such scattering event seems to violate the time-reversal symmetry. However, at this point of the Fermi contour the two spin-split bands are close to be degenerate. Therefore both spin partners are present, allowing the backscattering process, in a similar way as it happens on a Au(111) surface \cite{Petersen:2000}. Since backscattering is a preferred scattering process on metal surfaces, this feature has a rather high intensity in the FFT \cite{Petersen:2000}. The AC-ARPES map in Fig.\,\ref{fig:fig3}(b) does not reproduce it though, as the $1-\cos[\angle(\overrightarrow{k},\overrightarrow{k}+\overrightarrow{q})]$ term included in the integral excludes all the backscattering processes. %, not only $\overrightarrow{k} \rightarrow  -\overrightarrow{k}$ scattering.
This assignment of $q_2$ is further confirmed by the bias evolution
of this feature (Fig.\,\ref{fig:fig3}(a)). As the energy increases,
the bands which cross the Fermi level along $\bar{\Gamma} \bar{X}_1$
direction disperse towards the $\bar{\Gamma}$ point. Thus, the
feature $q_2$ moves towards the center of the FT-STM image
(Fig.\,\ref{fig:fig3}(a)), until the top of the band is reached and
it disappears from the FFT. This point at $\sim 100 meV$ is not
accurately reproduced by the band structure calculations in Fig. 1a
($\sim 190 meV$), what we attribute to the inability of DFT to
reproduce all fine structure details in the meV range. The spots
marked as $q_3$ result from the forward scattering between the edges
of the butterfly wing (the electron pocket) and the $\bar{\Gamma}
\bar{X}_1$-crossing point. The analysis of the band's dispersion also
shows that these features should move toward the center of the FFT
image (Fig.\,\ref{fig:fig3}(c)).

The identification of the scattering events far above the Fermi level
($q_4$ and $q_5$) is not as clear, because of the uncertainties in the
calculated dispersion and the absence of photoemission data. We can
tentatively explain their origin by the appearance of the new high DOS
feature (peak at 100\,meV in Fig.\,\ref{fig:fig1ARPES}(b)), when the
sample bias reaches the top of the bands near the $\bar{M}$ and
$\bar{\Gamma}$ point.

It is interesting to compare these results to the previous study of
electron scattering on Bi(110), another extreme case of the BiSb alloy. In
Bi(110) the pronounced spin-splitting of the surface bands leads to a
clear manifestation of the spin-dependent scattering process. In fact,
only spin conserving scattering events are observed in the Fourier
transformed dI/dV maps on Bi(110) \cite{Pascual2004,Strozecka:2011}.

Antimony has a smaller spin-orbit coupling and the spin-split of the bands
is generally smaller \cite{Bianchi:2012}. Although, the bands maintain a
certain spin texture (as evidenced in Fig. 2), the role of the spin in the
formation of the interference patterns is experimentally less evident, and
certain backscattering events could be identified. These are caused
because at certain points of the Fermi surface with high DOS, the two
sub-bands are so close in K-space that scattering between them will
resemble the backscattering events in the interference patterns, in a
similar way as described in Ref. \cite{Petersen:2000}.

In summary, we have predicted theoretically and confirmed experimentally
the unconventional spin texture of the Sb(110) surface. In spite of a
smaller spin-orbit coupling, the spin polarization of the surface bands
still dominates the electron dynamics of this surface. The LDOS
interference patterns formed around single impurities reveal the dominance
of non-direct backscattering events as in the related compound Bi(110).
The identification of the scattering events  was done combining  ARPES
measurements at E$_F$ with the state's dispersion above E$_F$, measured
from  energy dependent differential conductance maps, and interpreted on
the basis of DFT calculations of the spin texture and dispersion of the
surface state bands. The smaller spin-orbit interaction of Sb(110) is
reflected in a smaller band splitting what, in some cases, give rise to
interband scattering with wavevectors close to the pure backscattering
case.

\ack The authors gratefully acknowledge the support by  the Deutsche
Forschungsgemeinschaft (STR 1151/1-1), the Spanish Ministry of Science and
Innovation (FIS2010-19609-C02-00) and the Danish Council for Independent
Research– Natural Sciences,.

\section*{References}
%\bibliographystyle{unsrt}
%\bibliography{Ref}

\begin{thebibliography}{99}

\bibitem{Agergaard2001} Agergaard A, S{\o}ndergaard Ch, Li H, Nielsen M B,
    Hoffmann S V, Li Z and Hofmann Ph 2001 \emph{New J. Phys.} \textbf{3}
    15

\bibitem{Hofmann2006} Hofmann Ph 2006 \emph{Prog. Surf. Sci. }\textbf{81}
    191

\bibitem{Bianchi:2012} Bianchi M, Guan D, Strozecka A, Voetmann C H, Bao
    S, Pascual J I, Eiguren A and Hofmann Ph 2012 \emph{Phys. Rev. B}
    \textbf{85} 155431

\bibitem{Sugawara:2006} Sugawara K, Sato T, Souma S, Takahashi T, Arai M
    and T Sasaki 2006 \emph{Phys. Rev. Lett.} \textbf{96} 046411

\bibitem{kadono:2008} Kadono T, Miyamoto K, Nishimura R, Kanomaru K, Qiao
    S, Shimada K, Namatame H, Kimura A and Taniguchi M 2008 \emph{Appl.
    Phys. Lett.} \textbf{93} 252107

\bibitem{Hsieh2008} Hsieh D, Qian D , Wray L, Xia Y, Hor Y S, Cava R J and
    Hasan M Z 2008 \emph{Nature} \textbf{452} 970

\bibitem{Guo:2011} Guo H, Sugawara K, Takayama A, Souma S, Sato T, Satoh
    N, Ohnishi A, Kitaura M,
 Sasaki M, Xue Q K and Takahashi T 2011 \emph{Phys. Rev. B} \textbf{83} 201104

\bibitem{Teo:2008} Teo J C Y, Liang Fu and  Kane C L 2008 \emph{Phys. Rev.
    B } \textbf{78} 045426

\bibitem{Hsieh2009b} Hsieh  D, Xia Y, Wray L, Qian D, Pal A, Dil J H,
    Osterwalder J, Meier F, Bihlmayer G, Kane C L, Hor Y S, Cava R J and
    Hasan M Z 2009 \emph{Science} \textbf{323} 919

\bibitem{Hsieh:2010} Hsieh D, Wray L, Qian D, Xia Y, Dil J H, Meier F,
    Patthey L, Osterwalder J, Bihlmayer G, Hor Y S, Cava R J and Hasan M Z
    2010 \emph{New J. Phys.} \textbf{12} 125001

\bibitem{Park:2011} Park S R, Jung W S, Han G R, Kim Y K, Kim Chul, Song D
    J, Koh Y Y, Kimura S, Lee K D, Hur N, Kim J Y, Cho B K, Kim J H, Kwon
    Y S, Han J H and Kim C 2011 \emph{New J. Phys.} \textbf{13} 013008

\bibitem{Roushan2009} Roushan P, Seo J, Parker C V, Hor Y S, Hsieh D, Qian
    D, Richardella A, Hasan M Z, Cava R J and Yazdani A 2009 \emph{Nature}
    \textbf{460} 1106

\bibitem{Gomes:arxiv} Gomes K K, Ko W, Mar W, Chen Y, Shen Z X and
    Manoharan H C 2009 \emph{arXiv:}0909.0921

\bibitem{Pascual2004} Pascual J, Bihlmayer G, Koroteev Y, Rust H P,
    Ceballos G, Hansmann M, Horn K, Chulkov E, Bl\"{u}gel S, Echenique P
    and Hofmann Ph 2004 \emph{Phys. Rev. Lett.} \textbf{93} 196802

\bibitem{Alpichshev2010} Alpichshev Z, Analytis J G, Chu J H, Fisher I R,
    Chen Y L, Shen Z X, Fang A and Kapitulnik A 2010 \emph{Phys. Rev.
    Lett.} \textbf{104} 016401

\bibitem{Zhang2009a} Zhang T, Cheng P, Chen X, Jia J F, Ma X, He K, Wang
    L, Zhang H, Dai X, Fang Z, Xie X and Xue Q K 2009 \emph{Phys. Rev.
    Lett.} \textbf{103}

\bibitem{Strozecka:2011} Strozecka A, Eiguren A and Pascual J I. 2011
    Phys. Rev. Lett. \textbf{107} 186805

\bibitem{Petersen:2000} Petersen L and Hedegard P 2000 \emph{Surf. Sci. }
    \textbf{459} 49

\bibitem{Petersen1998} Petersen L, Sprunger P T, Hofmann Ph, Lagsgaard E,
    Briner B G, Doering M, Rust H P, Bradshaw A M, Besenbacher F and
    Plummer E W 1998 \emph{Phys. Rev. B} \textbf{57} 6858

\bibitem{Hoffmann:2004} Hoffmann S V, Sondergaard C, Schultz C, Li Z and
    Hofmann Ph 2004 \emph{Nucl. Instr. and Meth. in Phys. Res. A}
    \textbf{523} 441

\bibitem{Corso:2005} Dal Corso A and Mosca Conte A 2005 \emph{Phys. Rev.
    B} \textbf{71} 115106

\bibitem{pwscf}[www.pwscf.org]

\bibitem{Perdew:1996} Perdew J P, Burke  K and Ernzerhof M 1996
    \emph{Phys. Rev. Lett.} \textbf{77} 3865

\bibitem{Liu:1995} Liu Y and Allen R E 1995 \emph{Phys. Rev. B}
    \textbf{52} 1566

\bibitem{note} Additional features as those close to the line connecting
    $\Gamma$ and $X_2$ are not studied here due to the absence of clear
    dispersion patterns, what makes their association with cer-tain
    scattering event more speculative.

\bibitem{Markiewicz:2004} Markiewicz R S 2004 \emph{Phys. Rev. B}
    \textbf{69} 214517

\bibitem{McElroy2006} McElroy K, Gweon G H, Zhou S, Graf J, Uchida S,
    Eisaki H, Takagi H, Sasagawa T, Lee D H and Lanzara A 2006 \emph{Phys.
    Rev. Lett.} \textbf{96} 067005



\end{thebibliography}

\end{document}